# Electronic Reporting Using SM2-Based Ring Signcryption

YU Huifang, JIE Jiaxing, LI Lei

School of Cyberspace Security, Xi'an University of Posts and Telecommunications,
Xi'an 710121, China

Abstarct

Electronic whistleblowing systems are widely used due to their efficiency and convenience. The key to designing such systems lies in protecting the identity privacy of whistleblowers, preventing malicious whistleblowing, and ensuring the confidentiality of whistleblowing information. To address these issues, a SM2 traceable ring signcryption scheme for electronic voting is proposed. This scheme combines the SM2 elliptic curve public key cryptography algorithm with the ring signature algorithm, enhancing the overall efficiency of the scheme while ensuring the autonomy and controllability of the core cryptographic algorithms. Security analysis demonstrates that the proposed scheme satisfies confidentiality, unforgeability, traceability, linkability, and deniability. Efficiency analysis shows that, compared to existing ring signature schemes, the proposed scheme exhibits significant efficiency advantages during the signature phase. The electronic whistleblowing system designed using the proposed scheme can track malicious whistleblowers while protecting user identity privacy, and ensures that the content of whistleblowing remains unknown to third parties.

## 1 Introduction

With the development of information technology and Internet, electronic reporting system has gradually become an important channel for social governance and public supervision. Through this system, the public can easily and quickly submit information related to corruption, environmental pollution, security risks and other aspects, so as to effectively play the power of mass supervision. However, the reporting behavior usually involves sensitive content, and informants are generally worried about the disclosure of their true identity. Therefore, how to give consideration to the privacy protection of informants and the authenticity of the reported content in electronic reporting has become an important issue to be solved.

The existing electronic reporting platforms often have contradictions in identity protection and accountability. On the one hand, if there is insufficient privacy protection, the informant may not dare to report for fear of retaliation, resulting in the system function can not be brought into full play. On the other hand, if the system only emphasizes anonymity and lacks a tracking mechanism, it may be used by criminals, malicious reports and false reports emerge in endlessly, damaging the credibility of the system. How to balance "privacy protection" and "accountability" is the core challenge in the design of electronic reporting system.

Ring signatures were proposed by Rivest et al. [1] in 2001. They are a signature mechanism that provides strong anonymity, allowing a member of a group to sign on behalf of the entire group without revealing the signer's identity during the signing process, thus enabling the signer to conceal their identity within the group. Due to their outstanding performance in anonymity, ring signatures have been applied in various fields such as electronic voting[2], anonymous auctions[3], and electronic complaints[4]. Especially in the emerging field of blockchain technology[5], ring signatures are widely used as an important tool for protecting identity privacy. Although ring signatures have a wide range of application scenarios, their completely anonymous nature makes it impossible for regulators to hold malicious signers accountable, making them difficult to apply in regulatory environments. Liu et al.[6] proposed the concept of linkable ring signatures in 2004. Based on ring signatures, it allows users to generate a tag associated with the signature using their private key. This tag is unforgeable, and the same signer generates the same linkable tag for different messages, achieving the linking of signatures generated by the same signer. Komano et al.[7] proposed the concept of deniable ring signatures, which, in addition to the basic characteristics of ring signatures, enables ring members to prove whether they are the true signers of a ring signature through a confirmation/denial algorithm. Fujisaki et al.[8] furtherproposed the concept of traceable ring signatures (TRS), in which ring members can only anonymously express their opinion once for each tag. If a member submits another signature with different opinions for the same tag, their identity will be immediately exposed. This aims to prevent signers from abusing anonymity, achieving conditional anonymity. The characteristics of traceable ring signatures well meet the requirements of electronic reporting systems for privacy protection and accountability.

However, ring signatures cannot guarantee that the transmitted message is known only by a specific individual. Based on ring signatures, Huang et al.[9] proposed the concept of Ring Signcryption in 2005, aiming to achieve both digital signature and encryption in a single operation, thereby reducing computational and communication overhead while ensuring confidentiality and authenticity. Guo et al.[10] and Du et al.[11] applied Ring Signcryption to the Internet of Vehicles, demonstrating its performance advantages. The characteristic of Ring Signcryption in simultaneously ensuring confidentiality and authenticity also well meets the requirements of electronic reporting platforms, allowing the reporting management center to authenticate and protect the identity of the reporter while preventing third parties from knowing the content of the reporting information, thus better protecting the entire reporting process.

With the acceleration of localization process, China has gradually put forward a series of localization standards in daily and commercial fields. In the field of cryptography, state secret sm2[12] is an elliptic curve public key cryptography algorithm completely developed by China, taking into account high security and short key. In recent years, many foreign designed cryptographic algorithms have frequent vulnerabilities. In order to ensure the security of China's digital assets, we should try to use domestic cryptographic algorithms to replace foreign cryptographic algorithms. Fan et al.[13] proposed ring signature and linkable ring signature scheme based on SM2 digital signature. Bao et al.[14] proposed a deniable ring signature based on SM2 digital signature. In order to ensure the security of the electronic reporting platform, the domestic cryptographic algorithm should also be used in the design of relevant algorithms.

To solve the above problems, this paper proposes a SM2 traceable ring signcryption scheme for electronic reporting. The scheme in this paper is based on the state secret SM2 and combined with ring signcryption. We prove that the scheme in this paper meets the requirements of correctness, confidentiality, unforgeability, traceability and defamation. It also shows that the scheme in this paper also has conditional anonymity and linkability. We use the scheme in this paper to build an electronic reporting system, which not only ensures the privacy of users' identity, but also avoids the problem that malicious reports cannot be tracked, and ensures that the content of the reported information will not be obtained by third parties.

## 2 Preliminaries

### 2.1 Elliptic curve cryptography theory

Given a finite field $F(q)$ with prime order $q$, the elliptic curve e is defined on $F(q)$: $E: y^2 = x^3 + ax + b \mod q$, where $a, b \in F(q)$ and $4a^3 + 27b^2 \neq 0$, $G$ is a base point of elliptic curve $E$, and $G$ is also the generator of the additive cyclic group $\mathbb{G}$ with prime order $p$.

Elliptic curve discrete logarithm problem (ECDLP): given the point $P$ on elliptic curve $E$ and $P = kG$, $k \in [1, p-1]$ is an integer. If only the points $P$ and $G$ on the elliptic curve are given, it is difficult to solve the integer $k$.

### 2.2 SM2 digital signature algorithm

SM2 digital signature algorithm consists of four parts: system initialization (setup), keygen (keygen), signature (sign) and verify (verify).

(1) System initialization (setup): the system selects the security parameter $\lambda$ and finite field $F_q$ to define the two elements $a, b \in F_q$ of the equation of elliptic curve $E(F_q)$, and $4a^3 + 27b^2 \neq 0$; The base point $G = (x_G, y_G)(G \neq O)$, where $x_G$, $y_G$ are two elements in $F_q$; And $G$ is the generator of the additive cyclic group $\mathbb{G}$ composed of points satisfying the elliptic curve equation, and its order is $p$. $H: \{0,1\}^* \to Z_q^*$ is the secure hash function selected from SM3 cryptographic hash algorithm. Expose the system parameters $sp \leftarrow \{q, F_q, E(F_q), a, b, G, \mathbb{G}, p, H\}$.

(2) Keygen: user a generates its own key pair $(d_A, P_A)$, where $P_A = [d_A]G = (x_A, y_A)$.

(3) Signature:

    a. User $A$ has a length of $entlen_A$ discernible identification of bits $ID_A$, remember $ENTL_A$ Is an integer $entlen_A$ two bytes converted. Use the method given by SM2 elliptic curve public key cryptosystem [12] to get the bit string $Z_A = H(ENTL_A||ID_A||a||b||x_G||y_G||x_A||y_A)$. Calculate $\overline{M} = Z_A||M$, $e = H(\overline{M})$ for the signed message $M$, and convert the data type of e to an integer according to the method given by SM2 elliptic curve public key cryptography algorithm [12];

    b. The user randomly selects $k \in [1, n-1]$, calculates the elliptic curve point $(x_1, y_1) = kG$, and converts the data type of $x_1$ to an integer according to the method given by SM2 elliptic curve public key cryptography algorithm [12]. Calculate $r = (e + x_1) \mod n$, and return $b$ if $r = 0$ or $r + k = n$;

    c. Calculate $s_A = ((1 + d_A)^{-1}(k_A - c_A d_A)) \mod n$, and return $b$ if $s = 0$;

    d. According to the method given by SM2 elliptic curve public key cryptography algorithm[12], the data types of $r$ and $s$ are converted to byte strings, and the

signature of message $M$ is $(r,s)$.

(4) Verify:

a. Check the received message $M'$ and signature $(r',s')$. The verifier verifies whether $r' \in [1, n-1]$ and $s' \in [1, n-1]$ are valid. If not, the verification fails;

b. Calculate $\overline{M'} = Z_A||M'$, $e' = H(\overline{M'})$, and convert the data type of $e'$ to an integer according to the method given by SM2 elliptic curve public key cryptography algorithm [12].

c. Convert the data types of $r'$ and $s'$ into integers according to the method given by SM2 elliptic curve public key cryptography algorithm [12], and calculate $t = (r' + s') \bmod n$. if $t = 0$, the verification fails.

d. Calculate the elliptic curve point $(x_1', y_1') = [s']G + [t]P_A$, convert the data type of $x_1'$ to an integer according to the method given by SM2 elliptic curve public key cryptosystem [12], and calculate $R = (e' + x_1') \bmod n$. if $R = r'$ is true, the verification passes; Otherwise, the verification fails.

## 2.3 Ring signcryption

Ring signature was proposed by Rivest et al. At the Asiacrypt 2001 conference in 2001. Its core idea is that given a group composed of several users, any member can generate a signature on behalf of the whole group without the authorization of other members. After receiving the ring signature, the verifier can only confirm that the signature is indeed from a member of the group, but cannot distinguish the identity of the specific signer.

Ring signature generally consists of the following four polynomial time algorithms:

(1) System initialization algorithm ($Setup(\lambda) \rightarrow params$): a probabilistic polynomial time (PPT) algorithm. The input is the security parameter $\lambda$ and the output is the system parameter $params$.

(2) Key generation algorithm ($KeyGen(\lambda, params) \rightarrow (PK, sk)$): a PPT algorithm, which inputs the security parameter $\lambda$ and system parameter $params$, and outputs the user's public key $PK$ and private key $sk$.

(3) Ring signature generation ($RSign(M, n, S, sk_\pi) \rightarrow \sigma$): a PPT algorithm, which inputs the message to be signed $M$ and the set s composed of $n$ member public key $S = \{PK_1, \ldots, PK_n\}$ and signer private key $sk_\pi (1 \leq \pi \leq n)$, output the signature value $\sigma$.

(4) Ring signature verification ($Rverify(M, S, \sigma) \rightarrow accept/reject$): a deterministic algorithm whose inputs are the signature message $M$, the public key set $S$ and the signature value $\sigma$. If $\sigma$ is a valid ring signature of $M$, it outputs $accept$, otherwise it outputs $reject$.

However, ring signature cannot guarantee that only a specific person knows the message. Therefore, on the basis of ring signature, Huang et al. [9] proposed the concept of ring signcryption in 2005. Its purpose is to realize digital signature and encryption at the same time in one operation, so as to ensure confidentiality and authentication while reducing the operation and communication costs. Therefore, the ring signcryption algorithm is similar to the ring signature algorithm, but it combines digital signature and encryption in the signcryption generation phase.

Ring signcryption introduces the anonymity of ring signature based on the above, so that the receiver of the message can confirm that the message is indeed sent by a ring member, but cannot further identify the specific identity. Compared with the traditional "signature before encryption" method, ring signcryption not only achieves confidentiality, anonymity, unforgeability and integrity, but also has higher efficiency. Therefore, ring signcryption has shown important application value in the fields of electronic voting, anonymous authentication, privacy protection transactions and Internet of vehicles communication.

## 3 Security model

This section first introduces the following oracle machine. We initialize the **List** as the public key list maintained by KGC, **Mlist** as the list of malicious signers corrupted by the attacker, and **GSet** as the query oracle machine $CH_b(\cdot,\cdot,\cdot)$ message - signcryption pair list. The symbol $\emptyset$ indicates an empty set. The attacker cannot query the message/signcryption in **GSet** using oracle $C/D(\cdot,\cdot,\cdot)$.

-Join oracle machine $(JO(\bot) \to PK)$: Enter the user identity $id_i$ to generate a public-private key pair $(PK_i, sk_i)$, and will $id_i$ add to the **List** and output the user's public key $PK_i$.

-Corrupt oracle machine $(CO(PK_i) \to sk_i)$: Enter user identity $id_i$ and public key $PK_i$, will $id_i$ add to **MList** and output the user's private key $sk_i$.

-Signature oracle machine $(SO(M, n, S, PK_a) \to \sigma)$: Enter the signature message $m$, user identity $id_i$, public key set $S = \{PK_1, PK_2, \ldots, PK_n\}$, returns a valid ring signature $\sigma$.

-Challenge the oracle machine $CH_b(id_0, id_1, m) \to \sigma$: Enter the user's identity $id_0$ and $id_1$, message $m$, randomly select a bit $b \in \{0,1\}$, generate a signature $\sigma$, and set $(\sigma, id_0, id_1, m)$ add to **GSet** and output the signature value $\sigma$.

-Confirm/deny oracle $C/D(id_i, m, \sigma)$: Enter user identity $id_i$, message $m$, signature $\sigma$, if the user $id_i \notin$ List\Mlist, and the message signature pair queried does not belong to **GSet**, and the confirmation/denial algorithm is executed.

**Definition 1 (Confidentiality and unforgeability)** The scheme in this paper needs to resist two types of attackers. The attacker $A_1$ is an external fraudulent user who cannot obtain the master key but can launch a public key replacement attack; The attacker $A_2$ is a malicious KGC that has a master key but cannot launch a public key replacement attack. The security model is based on the interactive game between Challenger $C$ and $A_1$ or $A_2^{[15]}$. $C$ generates the corresponding parameters according to the signcryption algorithm and responds to the specific query of the opponent. The game ends when the query of $A_1$ or $A_2$ stops or triggers a specific condition. Finally, determine whether $A_1$ or $A_2$ can win the game in polynomial time. If $A_1$ or $A_2$ cannot win the interactive game with a non negligible advantage in polynomial time, the scheme in this paper is said to meet the confidentiality or unforgeability.

**Definition 2 (Non defamation)** Non defamation means that opponent $A$ cannot generate the same signature label as the real signer's signature, which ensures that $A$ cannot frame the real signer non defamation is defined by the following game between simulator $S$ and opponent $A$:

(1) Simulator $S$ generates the system parameter $sp$ and sends it to enemy $A$.

(2) $A$ can query adaptively and join the oracle machine $JO$.

(3) $A$ will sign the message $M$, a set of $n$ users $S = \{PK_1, PK_2, \ldots, PK_n\}$ and the randomly selected $a \in \{1,2,\ldots n\}$ are sent to $S$, where the public key of user $a$ not

corrupted or not used as input in $SO$ query of signature query; $S$ use $PK_a$ Corresponding private key $sk_a$ execute the ring signcryption generation algorithm, and finally send the generated signature value $\sigma$ to $A$.

(4) $A$ adaptive query oracle machine $JO$、$CO$、$SO$.

(5) $A$ output is based on public key set $S^* = \{PK_1^*, PK_2^*, \ldots, PK_n^*\}$ signature value of message $M^*$.

$A$ wins the above game if:

(1) Signature $\sigma^*$ verification passed

(2) $\sigma^*$ is not the output of the signing oracle $SO$;

(3) The public keys in $S^*$ are the outputs added to the oracle machine $JO$;

(4) $PK_a$ not corrupted by the enemy;

(5) The signature value $\sigma$ and $\sigma^*$ links were verified successfully.

If the probability of winning the above game is negligible for opponent $A$ of any PPT, the ring signcryption scheme is said to meet the non defamation requirement.

**Definition 3 (Traceability)** Traceability refers to the fact that the real signer of a ring signature can be traced by a confirmation/denial algorithm through the following experiments $Adv_{\Sigma,A}^{Trace}$ between simulator $S$ and PPT enemy $A$ to formally define traceability:

(1) Simulator $S$ generates the system parameter $sp$ and sends it to enemy $A$.

(2) Simulator $S$ initializes the lists $List \leftarrow \emptyset$、$MList \leftarrow \emptyset$、$GSet \leftarrow \emptyset$.

(3) Enemy $A$ adaptability query oracles $JO(\cdot)$、$CO(\cdot)$、$SO(\cdot)$、$C/D(\cdot)$.

(4) Enemy $A$ outputs message $m$, public key list $S = \{PK_1, PK_2, \ldots, PK_n\}$ and signature $\sigma$.

(5) If the signature verification is passed, 0

(6) If the user $A_1$, $A_2$, …, $A_n$ can deny $(m, \sigma)$, return 1; Otherwise, return 0

Define the advantage of adversary $A$: $Adv_{\Sigma,A}^{Trace} = Pr[Exp_{\Sigma,A}^{Trace} = 1]$. If against any ppt opponent $A$, the advantage $Adv_{\Sigma,A}^{Trace}$ is negligible, then the traceable ring signcryption scheme $\Sigma$ is traceable.

## 3 Scheme system model

The system model of the scheme in this paper is shown in Figure 1. The model has six components: regulatory center, regulators, regulatory center blockchain, government complaint/reporting website, reporting people and (Interplanetary File System, IPFS) IPFS cluster. Among them, the system model of this scheme adopts the certificateless key generation mechanism, and the supervision center acts as the KGC, avoiding the complex management problems in the whole life cycle of traditional digital certificates. In the system initialization phase, the supervision center uses Shamir secret sharing algorithm, and at least t of the N supervisors participate in the generation of the system master key $s$, which improves the resistance of the system to the risk of single point failure and enhances the security and robustness of the system. Details are as follows:

1) After the whistleblower logs in to the government complaint website, fill in the relevant whistleblower information and click send. After that, the interaction between users and the supervision center will be automatically completed by the complaint website, which will use the resources of users' computers to complete the subsequent interaction.

2) The government complaint website initiated some private key applications to the regulatory center. After receiving the request from the government complaint website, the supervision center will generate and return the corresponding part of the private key. At the same time, the supervision center will randomly extract the public key of n-1 users from the database and return it to the government complaint website.

3) After receiving the return from the supervision center, the government complaint website will sign and encrypt the reported information and upload the final ciphertext to the blockchain of the supervision center.

4) After receiving the ciphertext, the blockchain of the regulatory center will reach a consensus on the ciphertext, and upload the agreed ciphertext to the IPFs cluster after the consensus is completed.

5) After receiving the consensus ciphertext, IPFs will return the storage address and address hash value of the ciphertext to the regulatory center blockchain.

6) When some supervisors of the supervision center want to know the users of the reported information, they will send a data request to the blockchain of the supervision center after obtaining the permission of at least t supervisors (including themselves). After receiving the data request, the regulatory center blockchain verifies the identity of the regulatory center. After authentication, return the data storage address and address hash value to the supervision center.

7) After receiving the data storage address and address hash value, the supervision center verifies the hash value of the address, and then sends a data request to the IPFs cluster.

8) IPFs will verify the identity of the supervision center after receiving the data request. After authentication, return the requested data to the supervision center.

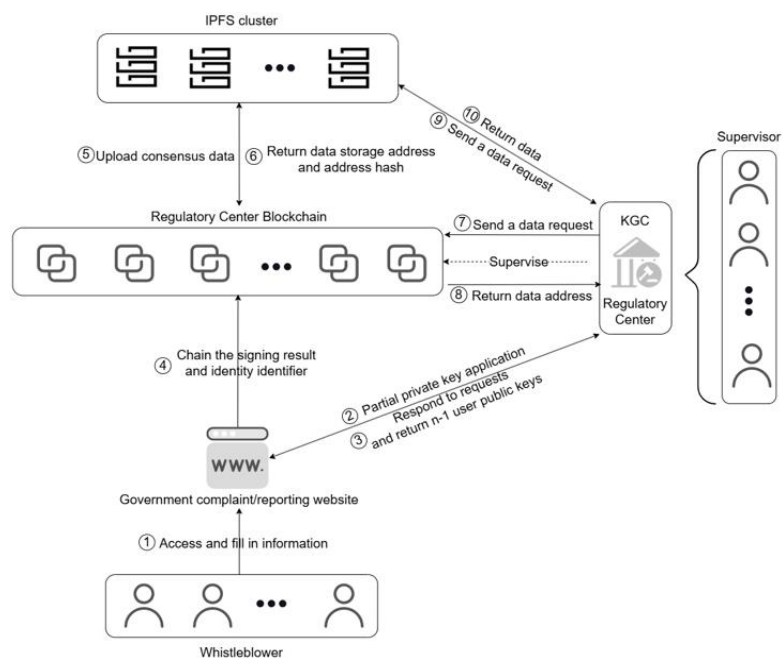

Figure.1: System model diagram based on this scheme

## 4 Traceable ring signcryption based on SM2

This section proposes a specific traceable ring signcryption scheme for SM2 for electronic reporting. The scheme consists of seven parts: system initialization, key generation, ring signcryption, de signcryption, signcrypter tracking, confirmation and denial. The specific description of each part is as follows:

## 4.1 System initialization algorithm

The supervision center selects five hash functions with cryptographic security strength:
$$H_0: \{0,1\}^* \times \mathbb{G} \times \mathbb{G} \to \mathbb{G}, \quad H_1: \mathbb{G}^{n_1} \to \mathbb{G}, \quad H_2: \{0,1\}^* \to Z_q^*$$
$$H_3: \mathbb{G} \times \mathbb{G} \times \mathbb{G} \times \mathbb{G} \times \mathbb{G} \times \mathbb{G} \times \{0,1\}^{n_2} \to Z_q^*$$
$$H_4: \mathbb{G} \times \mathbb{G} \times \mathbb{G} \times \mathbb{G} \times \mathbb{G} \times \mathbb{G} \to Z_q^*, \quad H_5: \times \mathbb{G}^n \times \mathbb{G} \times \mathbb{G} \to \{0,1\}^{n_2}$$

$n_1$ represents the number of public keys in the public key set, and $n_2$ represents the length of the message.

Supervisor center select the finite field $F_q$, select $a, b \in F_q$, and construct the elliptic curve $E: y^2 = x^3 + ax + b \bmod q$ over the finite field. $G$ is the base point of elliptic curve $E$ and the generator of additive cyclic group $\mathbb{G}$ of order $p$.

The supervision center generates $s \in Z_q^*$, public key $Y_{\text{pub}} = sG$. User identity set $U = \{ID_1, ID_2, \ldots, ID_n\}$.

Expose the system parameters $sp \leftarrow \{q, F_q, E, \mathbb{G}, G, p, Y_{\text{pub}}, H_0, H_1, H_2, H_3, H_4, H_5\}$.

The private key $s$ of the above supervision center shall be jointly generated by multiple supervisors in the supervision center through $(t, n)$ threshold algorithm. As follows: each supervisor $U_k$ generates a local polynomial:
$$f_k(x) = a_{k,0} + a_{k,1}x + a_{k,2}x^2 + \cdots + a_{k,t-2}x^{t-2} + a_{k,t-1}x^{t-1} \bmod q, \quad a_{k,w} \in Z_q^*$$

The constant term $a_{k,0}$ is the "personal contribution secret" of the participant.

Each supervisor $U_k$ sends shares to the $i$th Supervisor:
$$s_{k,i} = f_k(i) \bmod q$$

The $i$th supervisor collects all shares and sums them to get the global share:
$$s_i = \sum_{k=1}^{n} s_{k,i} \bmod q$$

The final global polynomial is:
$$f(x) = \sum_{k=1}^{n} f_k(x)$$

After a supervisor collected $t$ global shares $s_i$, he reconstructed the value of the global polynomial $f(x)$ at $x = 0$ through Lagrange interpolation, so the global secret is:
$$s = f(0) = \sum_{k=1}^{n} a_{k,0}$$

Verification:

To prevent a supervisor from $U_k$ sending inconsistent shares (cheating) to different people, verifiable secret sharing (VSS) is required:

Commitment of each supervisor $U_k$ to publish his polynomial coefficient:
$$C_{k,w} = a_{k,w}G, \quad w = 0,1,\ldots,t-1$$

After the receiver $U_k$ receives $s_{k,i}$, verify that:

$$s_{k,i}G = \sum_{w=1}^{t-1} i^w C_{k,w}$$

If it is true, it indicates that $s_{k,i}$ is correct; Otherwise, $U_k$ has cheated.

### 4.2 Key generation

Ring signer $ID_A$ select the secret value $x_A \in Z_q^*$ and calculate $X_A = x_A G$. Other users select the secret value $d_i \in Z_q^*$ and calculate the public key $P_i = d_i G$, where $i \neq A$ and $i \neq B$.

The signcryer sends the identity information through the secure channel $ID_A$ send it to the supervision center, and the supervision center randomly selects $\alpha \in Z_q^*$ to calculate the user's partial private key $\varphi_A = H_0(ID_A||Y_{\text{pub}}||\alpha G)$, $z_A = (y + \alpha)\varphi_A mod q$. The supervision center sends part of the private key $z_A$ to the ring signer $ID_A$. Ring signer $ID_A$ randomly select $\beta \in Z_q^*$ and calculate the complete private key $d_A: (x_A+\beta)z_A = (x_\beta, y_\beta)$, $d_A = x_\beta$, and then the signcryer calculates the complete public key $P_A = d_A G$.

The ring signcrypter arbitrarily selects the public keys of $n-1$ users and adds its own public keys to form the public key set $L = \{P_1, ..., P_n\} = \{(x_1, y_1), ..., (x_n, y_n)\}$, it is chained, and the relationship between user identity and public key $\{ID_i, P_i\}$ up chain. The ring signcryption is the $A$th user.

### 4.3 Ring signcryption

Signer randomly selects $h \in Z_q^*$ and calculates the random point $H = hG = (x_H, y_H)$. Calculate $E = H_5(ID_A||H||U||L||Y_{pub}) \oplus M$, where m is the message plaintext and the length is $n_2$.

Calculate $R = H_1(L)$, and link $Q_A = d_A R$. Construct identity $\Omega$: the ring signcryer randomly selects $r \in Z_q^*$, calculates $C_1 = rG$, $C_2 = P_A + rY_{\text{pub}}$, and makes $\Omega = (C_1, C_2)$.

The signer selects $k_A = H_2(h||Q_A)$ and calculates $c_{A+1} = H_3(C_1||C_2||P_A||Q_A||k_A(G+R)||E||Y_{\text{pub}})$.

For the index value, select $s_i \in Z_q^*$ randomly in the order of $A+1, ..., n, 1, ..., A-1$, and calculate:
$$Z_i = s_i(P_i + Q_A + G + R) + c_i(P_i + Q_A)$$
$$c_{i+1} = H_3(C_1||C_2||P_i||Q_i||Z_i||E||Y_{\text{pub}})$$

where $c_{n+1} = c_1$.

Calculate $s_A = (1 + d_A)^{-1}(k_A - c_A d_A)$. Ciphertext $\sigma = (Q_A, \Omega, E, c_1, s_1, ..., s_n)$ of output message $M$.

### 4.4 Designcryption

Only when the signer's identity is known can the message plaintext be recovered, and only the supervision center can know the signer's identity through signer tracking. After the regulatory center obtains the signer identity, calculate $M = H_5(ID_A||H||U||L||Y_{pub}) \oplus E$。

The regulatory center obtains the ciphertext $\sigma$, parses the ciphertext to get $Q_A$, $\Omega$, $c_1$, $s_1$, ..., $s_n$. Verify whether $c_1, s_1, ..., s_n \in Z_q^*$ is true. If not, the verification fails; Otherwise, proceed to the next step of verification: for all indexes $i \in Z_q^*$, calculate in the order of $1, ..., n$:

$$Z_i = s_i(P_i + Q_A + G + R) + c_i(P_i + Q_A)$$
$$c_{i+1} = H_3(C_1||C_2||P_i||Q_i||Z_i||E||Y_{\text{pub}})$$

when $c_{n+1} = c_1$, accept the recovered plaintext; Otherwise, the ciphertext will be rejected.

### 4.5 Signer tracking

When a supervisor or some supervisors of the supervision center want to know the identity of the signer, after the consent of at least $t$ supervisors (including themselves) (the $t$ supervisors' label is $i \in F$), the signer's public key can be obtained by the following methods:

In order to ensure that the private key $s$ of the supervision center will not be disclosed, the private key of the supervision center will not be completely recovered in the process of obtaining the signer's public key.

$t$ supervisors calculated and released partial decryption share $D_i = s_i C_1$, and made relevant zero knowledge proof for their own $s_i$.

Aggregate the decryption shares of the part, collect at least $t$ that have passed the verification $(i, D_i)$, calculate the Lagrange coefficient $\lambda$ (the same coefficient as the system initializes to calculate the private key $s$ of the supervision center, based on the index set of the supervisors involved in generating the private key), and then aggregate:

$$S = \sum_{i \in F} \lambda_i D_i = \sum_{i \in F} \lambda_i (s_i C_1) = (\sum_{i \in F} \lambda_i s_i) C_1 = sC_1$$

Then, according to the identity $\Omega$, calculate $C_2 - sC_1 = P'$, where $P'$ is the signer's public key. Then, according to the relationship between the user identity and the public key on the chain, the supervision center can know the signer's identity, so as to realize signer tracking.

### 4.6 Confirmation

The confirmation algorithm is performed between the prover and the verifier. The prover needs to prove that the ring signcryption is generated by himself. All users can run the confirmation algorithm:

The certifier randomly selects $r \in Z_q^*$, $P = rG$, $Q = rR$, $\eta = H_4(G, R, Q_A, P_A, P, Q)$. Calculate $e = r + \eta d_i \bmod q$ and send $\{e, \eta, P, Q\}$ to the verifier.

The verifier verifies whether $eG = P + cP_A$, $eR = Q + \eta Q_A$ is true. If it is true, it proves that the signcryption is indeed signed by the certifier.

### 4.7 Deny

The denial algorithm is performed between the prover (non signer, $i \neq A$) and the verifier. The prover needs to prove that the ring signature is not generated:

The certifier (user $i$) uses his private key $d_i$ to calculate his real link label $Q_i = d_i R$, and calculates the difference between the real label and the label in the signature: $D = Q_A - Q_i = Q_A - d_i R$.

The certifier sends the calculated difference point $D$ to the verifier, and then the certifier and the verifier interact as follows:

(a) The certifier randomly selects $k \in Z_q^*$ and calculates $T_1 = kG$, $T_2 = kR$. Send $(T_1, T_2)$

to the verifier.

(b) After receiving $(T_1, T_2)$, the verifier randomly selects $c \in Z_q^*$ and sends it to the certifier.

(c) After the certifier receives $c$, calculate $s = k + cd_i \bmod q$. The certifier sends $s$ to the verifier.

(d) After receiving $s$, the verifier verifies $sG = T_1 + cP_i$, $sR = T_2 + c(Q_A - D)$.

When the verifier verifies that both equations pass, the verifier believes that the verifier is a non signer.

## 5 Security analysis

**Theorem 1** Correctness of SM2 traceable ring signcryption.

**Proof:** Suppose the SM2 traceable ring signcryption value $\sigma = (Q_A, \Omega, E, c_1, s_1, \ldots, s_n)$ of the generated message $M$. The correctness of verification algorithm, confirmation algorithm, denial algorithm and tracking algorithm are discussed respectively:

(1) The correctness of the verify algorithm. The verifier verifies the signature through the verification algorithm. First, it is easy to know $c_1, s_1, \ldots, s_n \in Z_q^*$.

When verifying, $i$ increases from 1 to $n$, and the verifier calculates $Z_i = s_i(P_i + Q_A + G + R) + c_i(P_i + Q_A)$, $c_{i+1} = H_3(C_1 || C_2 || P_i || Q_i || Z_i || E || Y_{\text{pub}})$. When $i \neq A$, there is obviously $c_{i+1} = H_3(C_1 || C_2 || P_i || Q_i || Z_i || E || \theta || Y_{\text{pub}})$; When $i = A$, there are:

$$Z_A = s_A(P_A + Q_A + G + R) + c_A(P_A + Q_A)$$
$$= (1 + d_A)^{-1}(k_A - c_A d_A)(d_A G + d_A R + G + R) + c_A(d_A G + d_A R)$$
$$= (1 + d_A)^{-1}(k_A - c_A d_A)((1 + d_A)(G + R)) + c_A(d_A G + d_A R)$$
$$= (k_A - c_A d_A)(G + R) + c_A(d_A G + d_A R)$$
$$= k_A(G + R)$$

$$c_{A+1} = H_3(C_1 || C_2 || P_i || Q_i || Z_A || E || Y_{\text{pub}})$$
$$= H_3(C_1 || C_2 || P_i || Q_i || k_A(G + R) || E || Y_{\text{pub}})$$

Therefore, $c_1 = c_{n+1}$ is satisfied, so the correctness of the verification algorithm is satisfied.

(2) The correctness of the confirm algorithm. The signer proves that he has signed the message $M$ through the confirmation algorithm.

Because $Q_A = d_A R$, $P_A = d_A G$, $P = rG$, $Q = rR$, $e = r + cd_i \bmod q$. So:

$$P + cP_A = rG + cd_A G = (r + cd_A)G = eG$$
$$Q + cQ_A = rR + cd_A R = (r + cd_A)R = eR$$

Only when it is verified that this is the signer, can there be $d_i = d_A$, which makes the above formula true. The non signer can only guess the value of $d_i$ to make this formula true, while $d_i \in Z_q^*$, that is, the probability that the non signer has signed the message $M$ through the confirmation algorithm is $1/q$, which can be ignored.

(3) The correctness of the deny algorithm. The non signer $(i \neq A)$ proves that he has not signed the message $M$ through the denial algorithm. Among $Q_i = d_i R$, $D = Q_A - Q_i = Q_A - d_i R$, $Q_A = d_A R$, $P_A = d_A G$, $T_1 = kG$, $T_2 = kR$, $s = k + cd_i \bmod q$.

After receiving the difference point $D$ from the honest certifier, the verifier first checks whether $D$ is the zero point. If $D$ is the zero point, it indicates that the certifier is the signcryption person, and the verifier determines that denial failed; Otherwise, proceed to the next step of verification:

When $Q_A \neq Q_i$, calculate:

$$sG = (k + cd_i)G = kG + cd_iG = T_1 + cP_i$$
$$sR = (k + cd_i)R = kR + cd_iR = T_2 + cQ_i = T_2 + c(Q_A - D)$$

That is, the verifier successfully proves that he is not a signer.

(4) Correctness of tracking algorithm

The supervision center obtains the signcryption identity through the tracking algorithm, so as to realize the tracking of signcryption. Because $C_1 = rG$, $C_2 = P_A + rY_{pub}$, there is $C_2 - sC_1 = P_A + rY_{pub} - rsG = P_A + rY_{pub} - rY_{pub} = P_A$. The signcryption public key can be successfully recovered, so the tracking algorithm of SM2 traceable scheme is correct.

**Theorem 2** If the probabilistic polynomial time adversary $A_1$ can break the IND-CCA2-I security of TRSE-SM2 with the advantage of $\varepsilon$, there must be a challenge algorithm $C$ can solve the ECDLP problem.

It is proved that challenger $C$ receives a random instance of the ECDLP problem $(G, V = sG)$. The goal of $C$ is to calculate $V = sG \in \mathbb{G}$ using the external opponent $A_1$ playing the subroutine. $s \in Z_q^*$ is unknown to $C$. (list0, list1, list2, list3, list4, list5, list6) record the query and response values of $H_i (i = 0,1,2,3,4,5)$ oracle and public key oracle. These lists were initially empty. $C$ selects an integer $\tau$ from $\{1, 2, \ldots, l_0\}$, $l_0$ represents the number of times to query the $H_0$ oracle, $ID_\tau$ indicates the identity of the challenger, $\delta$ indicates the probability of $ID_i = ID_\tau$, $(\tau, ID_\tau)$ is unknown for $A_1$.

At the beginning of the game, $C$ runs the initialization algorithm to get the system global parameter $sp(Y_{pub} = sG \in \mathbb{G})$, and outputs $sp$ to $sp给A_1$. In stage 1, $A_1$ sends a series of polynomial bounded adaptive queries to $C$.

Public key query: $A_1$ query for $ID_i$ public key for. If list6 exists $ID_i$ the public key $P_i$, $C$ outputs $P_i$ to $A_1$; Otherwise, $C$ selects any $d_i \in Z_q^*$ and outputs the calculated $ID_i$ public key $P_i = d_iG \in \mathbb{G}$, and then store $(ID_i, d_i, P_i, -)$ to list6.

$H_0$ query: $A_1$ sends a message for $ID_i$ for $H_0$ query. $C$ checks whether there are matching tuples in list0. If it exists, $C$ outputs $H_i$ to $A_1$; Otherwise, $C$ responds as follows:

Scenario 1: if $ID_i = ID_\tau$, $C$ set $\varphi_i = V$, output $\varphi_i$ to $A_1$, store $(ID_i, \varphi_i, -)$ to list0;

Scenario 2: if $ID_i \neq ID_\tau$, $C$ select $\lambda \in Z_q^*$ arbitrarily, calculate $\varphi_i = \lambda G$, output $\varphi_i$ to $A_1$, store $(ID_i, \varphi_i, \lambda)$ to list0.

$H_1$ inquiry: $A_1$ can ask the $H_1$ oracle at any time. $C$ check whether there are matching tuples in list1. If yes, $C$ outputs $R$ to $A_1$; Otherwise, the $C$ output randomly selects $R \in \mathbb{G}$ to $A_1$ and stores $(ID_i, L, R)$ to list1.

$H_2$ inquiry: $A_1$ can ask the $H_2$ oracle at any time. $C$ checks whether there are matching tuples in list2. If yes, $C$ outputs $k_i$ to $A_1$; Otherwise, $C$ outputs $k_i \in Z_q^*$ to $A_1$ and stores $(ID_i, k_i, Q_A)$ to list2.

$H_3$ inquiry: $A_1$ can ask the $H_3$ oracle at any time. $C$ check whether there are matching tuples in list3. If yes, $C$ outputs $c_i$ to $A_1$; Otherwise, $C$ output randomly selects $c_i \in Z_q^*$ to $A_1$ and stores $(ID_i, C_1, C_2, P_i, Q_i, Z_i, E, Y_{pub})$ to list3.

$H_4$ inquiry: $A_1$ can ask the $H_4$ oracle at any time. $C$ check whether there are matching tuples in list4. If yes, $C$ outputs $\eta_i$ to $A_1$; Otherwise, $C$ outputs arbitrary $\eta_i \in Z_q^*$ to $A_1$ for storage $(ID_i, G, R, P_A, Q_A, P, Q)$ to list4.

$H_5$ inquiry: $A_1$ can ask the $H_5$ oracle at any time. $C$ check whether there are matching tuples in list5. If yes, $C$ outputs $f$ to $A_1$; Otherwise, $C$ outputs $f \leftarrow \{0,1\}^{n_2}$ to $A_1$ and stores $(ID_i, U, L, Y_{pub}, H)$ to list5.

Partial private key query: $A_1$ query $ID_i$ partial private key. If this is the $\tau$ inquiry, $C$ terminates the game; Otherwise, $C$ calls $H_0$ oracle to get $\lambda \in Z_q^*$, outputs part of the private key $z_i \leftarrow \lambda Y_{\text{pub}}$ to $A_1$, and then uses $(ID_i, d_i, P_i, z_i)$ update the $(ID_i, d_i, P_i, -)$.

Private key query: $A_1$ query $ID_i$ private key. If this is the $\tau$ inquiry, $C$ terminates the game; Otherwise, $C$ outputs the private key $d_i \in Z_q^*$ obtained from list6 to $A_1$.

Replace public key: $A_1$ select $P_i'$ to replace $ID_i$ public key. If this is the $\tau$ inquiry, $C$ terminates the game; Otherwise, $C$ uses $(ID_i, -, P_i, z_i)$ replace $(ID_i, d_i, P_i, z_i)$ in list6.

Ring signcryption query: $A_1$ query for $(ID_a, L, m)$, $ID_b \in U$. If $ID_b \neq ID_\tau$, $C$ output the ciphertext $\sigma = (Q_a, \Omega, E, c_1, s_1, \ldots, s_n)$ obtained by running the ring signcryption algorithm to $A_1$. Otherwise, $C$ selects any $h_i \in Z_q^*$, calculate $H = h_i G$, $E = H_5(ID_a||H||U||L||Y_{\text{pub}}) \oplus m$. $C$ storage list $(ID_a, U, L, Y_{\text{pub}}, H)$ to list5.

For $i \in \{1,2,\ldots,n\} \cup \{i \neq b\}$, $C$ selects any $s_i \in Z_q^*$, $c_{b+1} = H_3(C_1||C_2||P_b||Q_b||k_b(G+R)||E_b||Y_{\text{pub}})$, according to index $b+1,\ldots,n,1,\ldots,b-1$ calculate:
$$Z_i = s_i(P_i + Q_b + G + R) + c_i(P_i + Q_b)$$
$$c_{i+1} = H_3(C_1||C_2||P_i||Q_i||Z_i||E||Y_{\text{pub}})$$

$C$ storage $(ID_i, C_1, C_2, P_i, Q_i, Z_i, E, Y_{\text{pub}})$ to list3.

For $i = b$, $C$ calculate as:
$$s_b = (1 + d_b)^{-1}(k_b - c_b d_b)$$
$$Z_b = s_b(P_b + Q_b + G + R) + c_{b-1}(P_b + Q_b)$$
$$c_{b+1} = H_3(C_1||C_2||P_b||Q_b||Z_b||E||Y_{\text{pub}})$$

$C$ output $\sigma = (Q_b, \Omega, E, c_1, s_1, \ldots, s_n)$ to $A_1$.

Decryption query: $A_1$ sends a message for $(ID_a, L, \sigma = (Q_b, \Omega, E, c_1, s_1, \ldots, s_n))$. If $ID_a \neq ID_\tau$, $C$ output the results of the actual operation of the de signcryption algorithm to $A_1$. Otherwise, $C$ searches the list5 for tuple $(ID_i, U, L, Y_{\text{pub}}, H)$, make the external enemy $A_1$ ask $(Y_{\text{pub}}, r, U)$ when the oracle $\mathcal{O}_{ECDLP}$ output 1. If this is the case, $C$ obtain the secret $x_a$ from $ID_a$ through $A_1$ or list5, calculate as:
$$P_a = x_a G$$
$$m = f \leftarrow H_5(ID_b||H||U||L||P_a) \oplus E$$
output $m$ to $A_1$.

In the challenge phase, $A_1$ asks $C$ about the challenge identity $(ID_a', ID_b' \in U')$ and equal length message $(m_0, m_1)$. Before the challenge, $A_1$ cannot ask $ID_a'$ the secret value and partial private key of prime. If $ID_a' \neq ID_\tau$, $C$ terminate the game; Otherwise, $C$ selects any $t \in \{0,1\}$, $V \in \mathbb{G}$, sets $H = hG \in \mathbb{G}$, and calculates:
$$P_a' = x_a' G$$
$$E' = f' \leftarrow H_5(ID_b'||H'||U'||L'||P_a') \oplus m_t$$

$C$ storage list $(ID_b', H', U', L', P_a')$ to list5.

For $i \in \{1,2,\ldots,n\} \cup \{i \neq b\}$, $C$ selects any $s_i' \in Z_q^*$, $c_{b+1}' = H_3(C_1'||C_2'||P_b'||Q_b'||k_b'(G+R)||E_b'||Y_{pub})$, calculated according to the index $b+1,\ldots,n,1,\ldots,b-1$:
$$Z_i' = s_i'(P_i' + Q_b' + G + R) + c_i'(P_i' + Q_b')$$
$$c_{i+1}' = H_3(C_1'||C_2'||P_i'||Q_i'||Z_i'||E'||Y_{pub})$$

C storage $(ID_i', C_1', C_2', P_i', Q_i', Z_i', E', Y_{pub})$ to list3.

For $i = b$, $C$ is calculated as:
$$s_b' = (1 + d_b')^{-1}(k_b' - c_b' d_b')$$
$$Z_b' = s_b'(P_b' + Q_b' + G + R) + c_b'(P_b' + Q_b')$$

$$c_{b+1}' = H_3(C_1'||C_2'||P_{bb}'||Q_b'||Z_b'||E'||Y_{pub})$$

output $\sigma' = (Q_b', \Omega', E', c_1', s_1', \ldots, s_n')$ to $A_1$.

In stage 2, $A_1$ repeatedly sends a polynomial bounded adaptability query to $C$, as in stage 1. During inquiry, $A_1$ cannot be extracted $ID_a'$ the secret value and part of the private key of prime, $A_1$ cannot be used for $\sigma' = (Q_b', \Omega', E', c_1', s_1', \ldots, s_n')$ ask about the declassification machine. If $t = t'$, $C$ outputs the solution of the ECDLP problem instance.

Probability analysis: During the query process of stage 1 or stage 2, the probability that $C$ will not terminate Game1 is $\delta^{l_p+l_s+l_r}$ [15]; The probability of not terminating Game1 when challenging is $(1-\delta)$, and the probability of $C$ not giving up the game is $\delta^{l_p+l_s+l_r}(1-\delta)$. This formula reaches the maximum value when $\beta = 1 - 1/(1 + l_p + l_s + l_r)$, then the probability of C not failing in Game1 is at least $1/e(l_p + l_s + l_r)$. In addition, $C$ then uniformly selects $V \in \mathbb{G}$ with a probability of $1/l_1$. Therefore, the probability of $C$ using external enemy $A_1$ to solve the ECDLP problem is:

$$\varepsilon' \geq \frac{\varepsilon}{el_1(l_p + l_s + l_r)}$$

where $l_1, l_p, l_s, l_r$ are the number of times $A_1$ asked $H_5$ oracle machine, partial private key oracle machine, private key oracle machine and public key replacement oracle machine respectively.

**Theorem 3** If the probabilistic polynomial time adversary $A_2$ can overcome the IND-CCA2-II security of TRSE-SM2 with the advantage of $\varepsilon$, there must be a challenge algorithm $C$ can solve the ECDLP problem.

It is proved that challenger $C$ receives a random instance of the ECDLP problem $(G, V = sG)$. The goal of $C$ is to calculate $V = sG \in \mathbb{G}$ using the internal opponent $A_2$ playing the subroutine. $s \in Z_q^*$ is unknown to $C$. (list0, list1, list2, list3, list4, list5, list6) record the query and response values of $H_i(i = 0,1,2,3,4,5)$ oracle and public key oracle. These lists were initially empty. $C$ selects an integer $\tau$ from $\{1,2,\ldots,l_0\}$, $l_0$ represents the number of times to query the $H_0$ oracle, $ID_\tau$ indicates the identity of the challenger, $\delta$ indicates $ID_i = ID_\tau$, $(\tau, ID_\tau)$ is unknown for $A_2$.

At the beginning of the game, $C$ runs the initialization algorithm to get the system global parameter $sp$ $(Y_{pub} = sG \in \mathbb{G})$, and outputs $(s, sp)$ to $A_2$. $A_2$ sends a series of polynomial bounded adaptive queries to $C$, which are exactly the same as phase 1 of theorem 2, but do not ask for part of the private key and replace the public key. $C$ also responds like theorem 2 phase 1.

At the end of the adaptability inquiry above, $A_2$ finally outputs a forged $\sigma' = (Q_b', \Omega', E', c_1', s_1', \ldots, s_n')$. During the inquiry, $A_2$ cannot be extracted $ID_b'$ the secret value of prime, $A_2$ cannot be used for $\sigma' = (Q_b', \Omega', E', c_1', s_1', \ldots, s_n')$ ask about the declassification machine. If $ID_b' = ID_\tau$, $C$ declares failure and terminates the game. Otherwise, another valid signature can be obtained according to the bifurcation $\sigma^* = (Q_b^*, \Omega^*, E^*, c_1^*, s_1^*, \ldots, s_n^*)$, $C$ Output the answers of ECDLP problem instances.

Probability analysis: during the query process of stage 1 or stage 2, the probability that $C$ will not terminate game2 is $\delta^{l_s}$ [15]; The probability of not terminating game2 during the challenge is $(1-\delta)$, and the probability of $C$ not giving up the game is $\delta^{l_s}(1-\delta)$, which reaches the maximum value when $\beta = 1 - 1/(1 + l_s)$, then the probability of $C$ not failing in game2 is at least $1/el_s$. In addition, $C$ then uniformly selects $V \in \mathbb{G}$ with a probability of $1/l_1$. Therefore, the probability for $C$ to solve the ECDLP problem by using its internal enemy $A_2$ is:

$$\varepsilon' \geq \frac{\varepsilon}{el_1 l_s}$$

where $l_1, l_s$ are the times $A_1$ asked $H_5$ oracle machine and private key oracle machine respectively.

**Theorem 4** If the probability polynomial time adversary $A_1$ can break the UF-CMA- I security of TRSE-SM2 with the advantage of $\varepsilon$, there must be a challenge algorithm $C$ that can solve the ECDLP problem.

It is proved that challenger $C$ receives a random instance of the ECDLP problem $(G, V = sG)$. The goal of $C$ is to calculate $V = sG \in \mathbb{G}$ using the external opponent $A_1$ playing the subroutine. $s \in Z_q^*$ is unknown to $C$.

At the beginning of the game, $C$ runs the initialization algorithm to get the system global parameter $sp(Y_{\text{pub}} = sG \in \mathbb{G})$, and outputs $sp$ to $A_1$. $A_1$ sends a series of polynomial bounded adaptive queries to $C$, and the queries are exactly the same as phase 1 of theorem 2. $C$ also responds like theorem 2 phase 1.

At the end of the adaptability inquiry above, $A_1$ finally outputs a forged ciphertext $\sigma' = (Q_b', \Omega', E', c_1', s_1', \ldots, s_n')$. During the inquiry, $A_1$ cannot be extracted $ID_b'$ the secret value and part of the private key of prime, $A_1$ cannot be used for $\sigma' = (Q_b', \Omega', E', c_1', s_1', \ldots, s_n')$ ask about the declassification machine. If $ID_b' = ID_\tau$, $C$ declares failure and terminates the game. Otherwise, another valid signature can be obtained according to the bifurcation $\sigma^* = (Q_b^*, \Omega^*, E^*, c_1^*, s_1^*, \ldots, s_n^*)$, $C$ output the answers of ECDLP problem instances.

Probability analysis: according to theorem 2, the probability of $C$ winning in Game1 is $1/e(l_p + l_s + l_r)$. Therefore, the probability of $C$ using external enemy $A_1$ to solve the ECDLP problem is:

$$\varepsilon' \geq \frac{\varepsilon}{e(l_p + l_s + l_r)}$$

where $l_p, l_s, l_r$ are the times that $A_1$ asked $H_5$ oracle machine, partial private key oracle machine, private key oracle machine and public key replacement oracle machine respectively.

**Theorem 5** If the probability polynomial time adversary $A_2$ can break the UF-CMA- II security of TRSE-SM2 with the advantage of $\varepsilon$, there must be a challenge algorithm $C$ that can solve the ECDLP problem.

It is proved that challenger $C$ receives a random instance of the ECDLP problem $(G, V = sG)$. The goal of $C$ is to calculate $V = sG \in \mathbb{G}$ using the internal opponent $A_2$ playing the subroutine. $s \in Z_q^*$ is unknown to $C$.

At the beginning of the game, $C$ runs the initialization algorithm to get the system global parameter $sp(Y_{\text{pub}} = sG \in \mathbb{G})$, and outputs $(s, sp)$ to $A_2$. $A_2$ sends a series of polynomial bounded adaptive queries to $C$. The queries are exactly the same as phase 1 of Theorem 3, but do not ask for partial private keys and replace public keys. $C$ also responds like theorem 2 phase 1.

At the end of the adaptability inquiry above, $A_2$ finally outputs a forged ciphertext

$\sigma' = (Q_b', \Omega', E', c_1', s_1', \ldots, s_n')$. During the inquiry, $A_2$ cannot be extracted $ID_b'$ the secret value of prime, $A_2$ cannot be used for $\sigma' = (Q_b', \Omega', E', c_1', s_1', \ldots, s_n')$ ask about the declassification machine. If $ID_b' = ID_\tau$, $C$ declares failure and terminates the game. Otherwise, another valid signature can be obtained according to the bifurcation $\sigma^* = (Q_b^*, \Omega^*, E^*, c_1^*, s_1^*, \ldots, s_n^*)$, $C$ Output the solution of ECDLP problem instance $\frac{s_b^* - s_b'}{x_a(s_b' - s_b^* + c_b' - c_b^*)}$.

Probability analysis: according to Theorem 3, the probability of $C$ winning in game2 is $1/el_s$. Therefore, the probability for $C$ to solve the ECDLP problem by using its internal enemy $A_2$ is:

$$\varepsilon' \geq \frac{\varepsilon}{el_s}$$

where $l_s$ is the number of times $A_1$ asked $H_5$ oracle and private key oracle.

**Theorem 6** SM2 traceable ring signcryption satisfies non defamation.

It is proved that given the example of ECDLP difficult problem, simulator $S$ generates the system parameter $sp$ and sends it to $A$; then $A$ can adaptively query and join the Oracle machine ($JO$); $A$ specifies the signature message $m$, the signature public key set $S^* = \{PK_1^*, PK_2^*, \ldots, PK_n^*\}$ and signer $a \in 1,2,\ldots n$ are sent to $S$, and the simulator generates the corresponding signature value $\sigma = (Q_a, \Omega, E, c_1, s_1, \ldots, s_n)$ and sends it to $A$; $A$ can adaptively query the oracle machine $JO$, $CO$, $SO$, but for $PK_a^*$ ask for corruption query; Finally, $A$ counterfeits another set of signature values $\sigma^* = (Q_a^*, \Omega^*, E^*, c_1^*, s_1^*, s_2^*, \ldots, s_n^*)$ and $Q_a^* = d_a^* H_1(S^*)$ of signer $a$, and it can pass the signature verification algorithm. Because the two signatures are interrelated, there is $Q_a^* = d_a^* H_1(S^*) = Q_a$, so the signer knows the private key $d_a^*$; Because $S$ uses the private key $d_a$ to generate $Q_a = d_a H_1(S^*)$, $d_a H_1(S^*) = d_a^* H_1(S^*)$, that is, $d_a d_a^{-1} H_1(S^*) = H_1(S^*)$, Thus $d_a = d_a^*$. Finally, it is concluded that $A$ has obtained the private key $d_a$, which is contradictory to that $A$ has not queried the private key $d_a$. Therefore, SM2 traceable ring signcryption satisfies non defamation.

**Theorem 7** SM2 traceable ring signcryption satisfies traceability.

**Proof:** We need to prove that there is no ppt enemy a can output an untraceable forged signature $\sigma$ with a non negligible probability. For all $i \in 1,2,\ldots,n$ is satisfied $P_i/G \neq Q_A/R$. Using the same method of proof to the contrary, it is assumed that adversary $A$ can forge such a signature. Through the bifurcation lemma [16], adversary $A$ can obtain two untraceable forged signatures, which is similar to the proof process of theorem 2. It can be found that there is an $i$, and the equation $P_i/G \neq Q_A/R$ holds, which is obviously inconsistent with the assumption that $Adv_{\Sigma,A}^{nf} = Pr[Exp_{\Sigma,A}^{nf} = 1]$ is negligible, that is, the scheme satisfies traceability.

## 6 Performance analysisi

### 6.1 Characteristic analysis

Conditional anonymity: the arbitrator accepted the ring signcryption $ID_A$ when the ciphertext $\sigma = (Q_A, \Omega, E, c_1, s_1, \ldots, s_n)$ is sent, it is resolved that $\Omega = (C_1, C_2)$. Because $C_1 = rG$, $C_2 = P_A + rY_{\text{pub}}$. According to the ECDLP problem, no user can recover $P_A$ from $C_2$ or know the relationship between $G$ and $Y_{\text{pub}}$. Because $P_A = C_2 - sC_1$, if you want to recover $P_A$, only if you know the arbitrator's master private key $s$ or $sC_1$. The arbitrator's master private key is distributed and stored according to the secret sharing algorithm, so only when multiple arbitrators cooperate can the arbitrator's master private key $s$ be obtained or $sC_1$ be recovered. It is easy to know that the signer is completely anonymous for users other than the arbiter; For the arbiter who recovers the master private key $s$ or $sC_1$, the signcrypter is non anonymous.

Linkability: when the arbitrator receives two ciphertexts $\sigma = (Q_A, \Omega, E, c_1, s_1, \ldots, s_n)$ and $\sigma' = (Q_A', \Omega', E', c_1', s_1', \ldots, s_n')$. Input the two sets of ciphertexts into the ring signcryption verification algorithm respectively. If both ciphertexts pass the verification, the arbitrator will further verify whether $Q_A = Q_A'$ is true. If it is true, it can be determined that the two ciphertexts are signed by the unified signer.

### 6.2 Performance optimization

The computational overhead of this scheme mainly focuses on the point multiplication on elliptic curves, and the point multiplication on elliptic curves mainly comes from $Z_i = s_i(P_i + Q_A + G + R) + c_i(P_i + Q_A)$ in ring signcryption cycles. Therefore, in order to reduce the computational overhead of this scheme, the Shamir/Straus double base joint multiplication method can be used to optimize the operation time.

Since the number $c_i$, $s_i$ in each round of $Z_i$ calculation in the loop generating ring signcryption needs to be output from the previous round, it is impossible to collect all scalars in advance for batch calculation during the signature generation process, so the pipenger/msm optimization algorithm cannot be directly used in this scheme. However, in order to verify the performance of pipenger/MSM optimization algorithm in large-scale multi scalar multiplication, we conducted experiments on typical multi scalar multiplication (MSM) scenarios under the same curve parameters.

Since the point multiplication on elliptic curves mainly comes from $Z_i = s_i(P_i + Q_A + G + R) + c_i(P_i + Q_A)$ in the ring signcryption cycle, we only optimize the time required for this cycle to run under different $n$. Table 1 and Figure 2 respectively show the time required for the loop without optimization, the loop with Shamir/Straus double base joint multiplication optimization and the loop optimized under the same curve parameter and the same curve parameter.

TABLE I TIME REQUIRED AFTER CYCLE OPTIMIZATION

|  | Not optimized | Shamir / Straus optimization | Pippenger / MSM optimization |
|---|---|---|---|
| n=10 | 1.1215s | 0.6658s | 0.2406s |
| n=30 | 3.3766s | 1.9790s | 0.5006s |
| n=50 | 5.6941s | 3.2837s | 0.6513s |
| n=70 | 7.8752s | 4.5616s | 0.8592s |
| n=90 | 10.1758s | 5.9262s | 1.0090s |

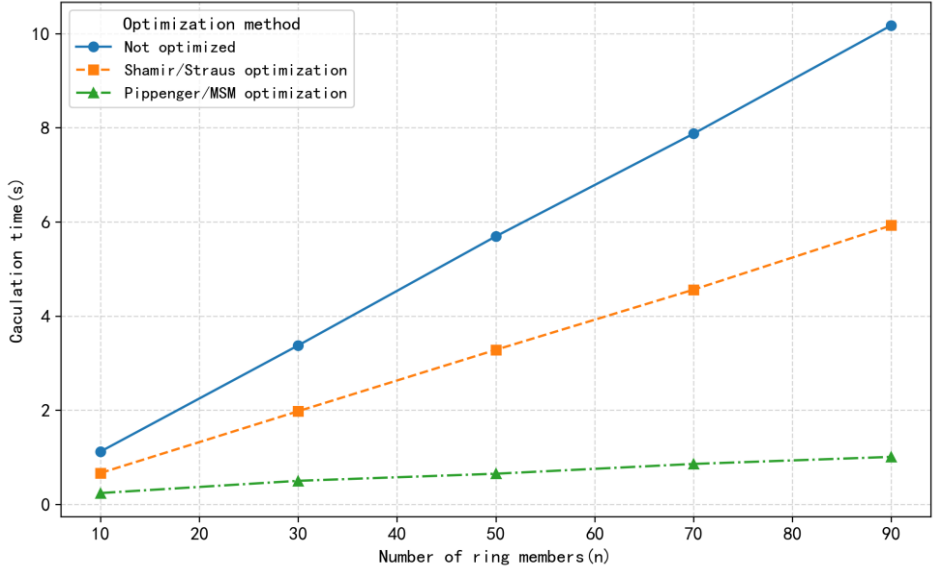

Fig. 1 Time required after cycle optimization

6.3 Calculation cost

In order to evaluate and compare the performance of our scheme with the other three schemes, we selected the elliptic curve secp256k1 on the finite field generated by 256bit large prime as the elliptic curve used in the experiment, and G is the base point of the elliptic curve. The experimental platform is configured as: Windows 11 operating system, AMD ryzen 7 4800h with radeon graphics, 16GB memory. The relevant algorithms of this scheme are completed in pychar using python programming language.

Table II lists the average time cost of some major password operations after running 1000 times. In addition, the element lengths generated by $Z_p^*$ and $G$ in this scheme are 256bits and 512bits respectively.

TABLE II AVERAGE TIME OF EACH OPERATION

| Symbol | Meaning | Cost |
|---|---|---|
| TPA | One addition operation | 0.0098 ms |
| TPM | One scalar multiplication | 2.7350 ms |
| TH | One hash operation | 0.0007 ms |
| THP | One string mapping to points operation | 0.1336 ms |
| TINV | One reverse operation | 0.0263 ms |

According to the average time required for the main password operations in Table 2, we evaluated the time cost of the scheme in this paper and the schemes in references [13], [14], [17] in several key stages, as shown in Table 2. We give the time cost comparison diagram of the scheme in this paper and those in references [13], [14], [17] in the signature generation phase and signature verification phase, as shown in figures 3, 4 and 5. At the same time, the time cost comparison diagram of SM2 repudiation ring signature scheme and our scheme in the confirmation phase and the denial phase is also given, as shown in Figure 6.

TABLE III TIME COST TABLE OF KEY STAGE

| Scheme | Signature generation | Signature verification | Confirm | Deny($v = 1$) |
|---|---|---|---|---|
| Literature [13] | $(2n-1)T_{PM} + (n-1)T_{PA} + nT_H + T_{INV}$ | $2nT_{PM} + nT_{PA} + nT_H$ | - | - |
| Literature [14] | $(4n-1)T_{PM} + (2n-2)T_{PA} + (n+1)T_H + T_{INV}$ | $4nT_{PM} + 2nT_{PA} + (n+1)T_H$ | $10T_{PM} + 4T_{PA}$ | $(10+z/2)T_{PM} + 6T_{PA} + 2T_H$ |
| Literature [17] | $(4n-2)T_{PM} + (2n-2)T_{PA} + 2T_H + (2n-1)T_{HP}$ | $4nT_{PM} + T_H$ | - | - |
| Our scheme | $2nT_{PM} + (5n-5)T_{PA} + (n+1)T_H + T_{INV}$ | $2nT_{PM} + 5nT_{PA} + (n+1)T_H$ | $6T_{PM} + 2T_{PA} + T_H$ | $7T_{PM} + 2T_{PA}$ |

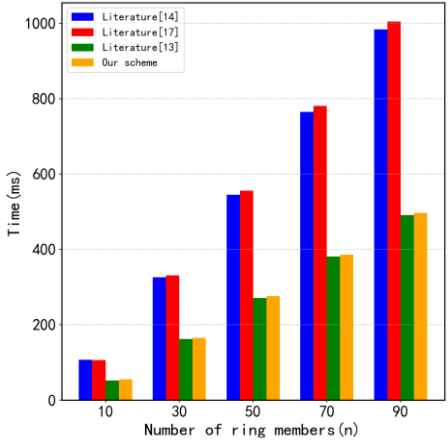 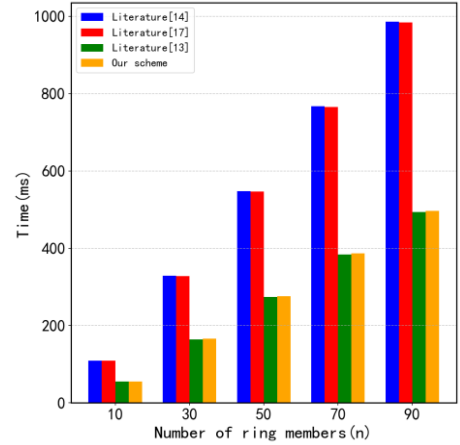

Fig. 3 Time cost in signature generation stage    Fig. 4 Time cost in signature verification phase

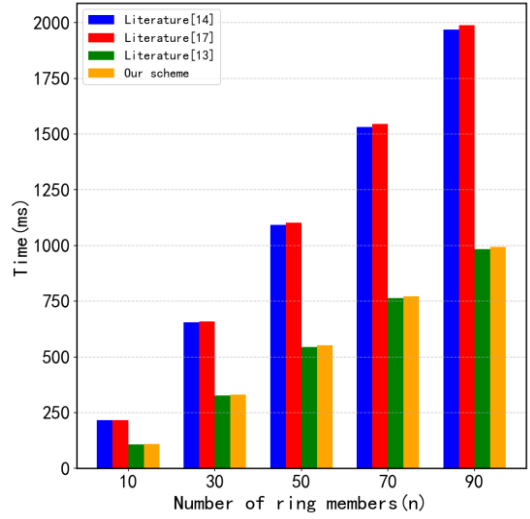

Fig. 5 Time cost in signature generation and verification stage

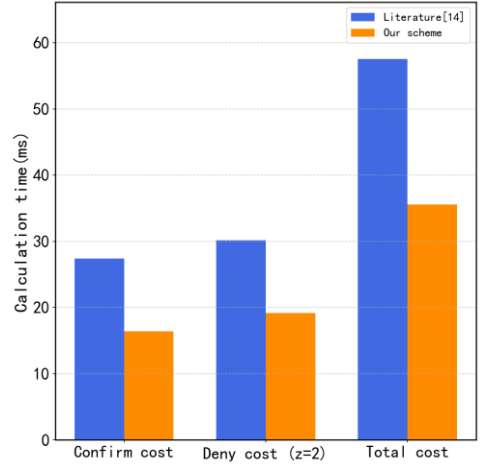

Fig. 6 Time cost in confirm and deny stage

6.4 Communication expenses

    The length of the signature is an important factor determining the communication cost of the signature scheme. In the experimental process of this article, SECP256K1 elliptic curve is used, where $|Z_p^*|$ and $|G|$ represent the length of an element in $Z_p^*$ and $G$, respectively. As shown in Table 4 and Figure 7, Figure 8 compares the communication overhead of SM2 deniable ring signature and our proposed scheme in terms of acknowledgment and denial algorithms.

TABLE IV COMPARISION TABLE OF COMMUNICATION COST

| Scheme | Signature algorithm | Confirm algorithm | Deny algorithm($v = 1$) |
|---|---|---|---|
| Literature [13] | $(n+1)\|Z_q^*\|$ | - | - |
| Literature [14] | $(n+2)\|Z_q^*\| + \|\mathbb{G}\|$ | $3\|Z_q^*\| + 3\|\mathbb{G}\|$ | $(4\|Z_q^*\| + \|z\| + 2\|\mathbb{G}\|)v$ |
| Literature [17] | $2n\|Z_q^*\| + \|\mathbb{G}\|$ | - | - |
| Our scheme | $(n+2)\|Z_q^*\| + 3\|\mathbb{G}\|$ | $2\|Z_q^*\| + 2\|\mathbb{G}\|$ | $2\|Z_q^*\| + 3\|\mathbb{G}\|$ |

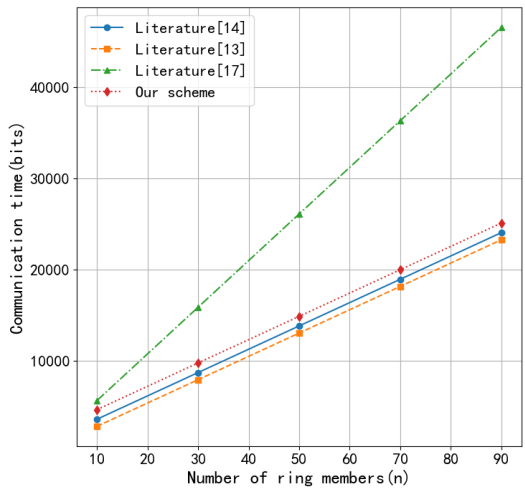
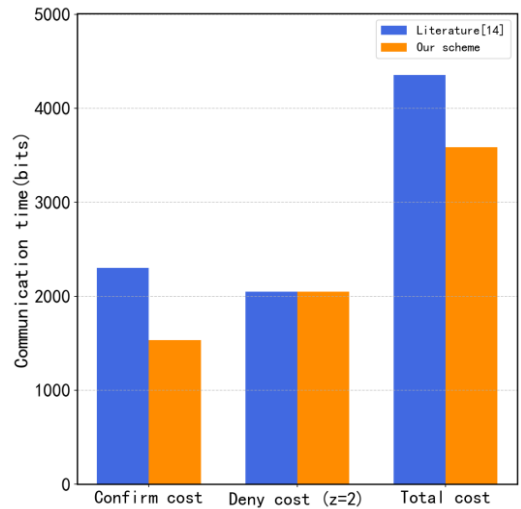

Fig. 7 Time cost in signature algorithm    Fig. 8 Time cost in confirm and deny algorithm

## 7 Conclusion

The SM2 traceable ring signature encryption scheme for electronic reporting not only ensures the identity privacy of the reporting public, but also enables the tracking of malicious reporting users. The proposed algorithm is based on the SM2 elliptic curve public key cryptography algorithm, which takes into account the privacy protection of whistleblowers, the authenticity of reported content, and the confidentiality of reported content while being localized. In future work, the team will focus on developing lightweight and secure solutions, and further explore their practical applications in network environments.

## References

[1]  RIVEST R L,SHAMIR A,TAUMAN Y. How to leak a secret[C]∥7th International Conference on the Theory  and Application of Cryptology and Information Security.  2001:552-565.

[2]  QIU C, ZHANG S B, CHANG Y, et al. Electronic voting scheme based on a quantum ring signature[J]. International Journal of Theoretical Physics, 2021, 60(4): 1550–1555. [DOI: 10.1007/s10773-021-04777-1]

[3]  YE J W, KANG X, LIANG Y C, et al. A trust-centric privacy-preserving blockchain for dynamic spectrum management in IoT networks[J]. IEEE Internet of Things Journal, 2022, 9(15): 13263–13278. [DOI: 10.1109/JIOT.2022.3142989]

[4]  Zhang Zisong, Zhang Bowen, Tong Xinhai. A BLOCKCHAIN ELECTRONIC COMPLAINT REPORTING SCHEME   BASED ON RING SIGNATURE[J]. Computer Applications and Software, 2024, 41(7): 329-335. DOI: 10.3969/j.issn.1000-386x.2024.07.047

[5]  VAN SABERHAGEN N. CryptoNote v2.0[EB/OL]. White Paper. 2013. https://bytecoin.org/old/whitepaper.pdf

[6]  LIU J K, WEI V K, WONG D S. Linkable spontaneous anonymous group signature for ad hoc groups[C]. In: Information Security and Privacy—ACISP 2004. Springer Berlin Heidelberg, 2004: 325–355. [DOI: 10.1007/9783-540-27800-9_28]

[7]  KOMANO Y, OHTA K, SHIMBO A, et al. Toward the fair anonymous signatures: Deniable

ring signatures[C]. In: Topics in Cryptology—CT-RSA 2006. Springer Berlin Heidelberg, 2006: 174–191. [DOI: 10.1007/11605805_12]

[8] FUJISAKI E, SUZUKI K. Traceable ring signature[C]//2007 International Workshop on Public Key Cryptography. Berlin: Springer, 2007: 181-200.

[9] Huang X, Susilo W, Mu Y, et al. Identity-based ring signcryption schemes: Cryptographic primitives for preserving privacy and authenticity in the ubiquitous world[C]//19th International Conference on Advanced Information Networking and Applications (AINA'05) Volume 1 (AINA papers). IEEE, 2005, 2: 649-654.

[10] Guo R, Xu L, Li X, et al. An efficient certificateless ring signcryption scheme with conditional privacy-preserving in VANETs[J]. Journal of Systems Architecture, 2022, 129: 102633.

[11] Du H, Wen Q, Zhang S, et al. An improved conditional privacy protection scheme based on ring signcryption for vanets[J]. IEEE Internet of Things Journal, 2023, 10(20): 17881-17892.

[12] State Cryptography Administration. Public Key Cryptographic Algorithm SM2 Based on Elliptic Curves[S/OL]. 2010.12. http://www.sca.gov.cn/sca/xwdt/2010-12/17/content_1002386.shtml

[13] FAN Qing, HE De-Biao, LUO Min, HUANG Xin-Yi, LI Da-Wei. Ring Signature Schemes Based on SM2 Digital Signature Algorithm. *Journal of Cryptologic Research*. 2021, 8(4): 710-723 https://doi.org/10.13868/j.cnki.jcr.000472

[14] Bao Z, He D, Peng C, et al. Deniable ring signature scheme based on sm2 digital signature algorithm[J]. Journal of Cryptologic Research, 2023, 10(2): 264-275.

[15] Yu H, Wang Z, Yang B. Decentralized Electronic Cash Payment Scheme Using SM2-Based Blind Signcryption[J]. IEEE Internet of Things Journal, 2025.

[16] POINTCHEVAL D, STERN J. Security arguments for digital signatures and blind signatures[J]. Journal of Cryptology, 2000, 13(3): 361–396. [DOI: 10.1007/s001450010003].

[17] WANG L, PENG C G, TAN W J. Secure ring signature scheme for privacy-preserving blockchain[J]. Entropy, 2023, 25(9):1334.